\documentclass[final]{vgtc}                          % final (conference style)
\graphicspath{{figures/}{pictures/}{images/}{./}} % where to search for the images

\usepackage{times}                     % we use Times as the main font
\usepackage{tabu}                      % only used for the table example
\usepackage{booktabs}                  % only used for the table example
\usepackage{lipsum}                    % used to generate placeholder text
\usepackage{mwe}                       % used to generate placeholder figures

\usepackage{mathptmx}                  % use matching math font

\usepackage{subcaption} % Added by iddo 
\usepackage{arydshln}

\onlineid{0}

\vgtccategory{Research}

\vgtcinsertpkg

\title{Tactile Search: Enhancing Targeting in 3D Space}

\author{
Amber Maimon\thanks{These authors contributed equally to this work.}\, \thanks{e-mail: maimon.amber@gmail.com}\\ %
\parbox{1.5in}{\scriptsize \centering University of Canterbury \\ Christchurch, New Zealand \\ University of Bremen \\ Bremen, Germany \\ University of Haifa \\ Haifa, Israel \\ Ben Gurion University \\ Be'er Sheva, Israel} \\[0.1em]
\and
Iddo Yehoshua Wald\footnotemark[1]\, \thanks{e-mail: wald@uni-bremen.de}\\ %
\parbox{1.5in}{\scriptsize \centering Digital Media Lab \\ University of Bremen \\ Bremen, Germany}
\and
Jonas Keppel\thanks{e-mail: jonas.keppel@uni-due.de}\\ %
\parbox{1.5in}{\scriptsize \centering University of Duisburg-Essen \\ Essen, Germany}
\and
Eunhee Chang\thanks{e-mail: chang.eunhee.d4@tohoku.ac.jp}\\ %
\parbox{1.5in}{\scriptsize \centering Tohoku University \\ Sendai, Japan}
\and
Yoshifumi Kitamura\thanks{e-mail: kitamura@riec.tohoku.ac.jp}\\ %
\parbox{1.5in}{\scriptsize \centering Tohoku University \\ Sendai, Japan}
\and
Stefan Schneegass\thanks{e-mail: stefan.schneegass@uni-due.de}\\ %
\parbox{1.5in}{\scriptsize \centering University of Duisburg-Essen \\ Essen, Germany} \\[0.3em]
\and
Rainer Malaka\thanks{e-mail: malaka@uni-bremen.de}\\ %
\parbox{1.5in}{\scriptsize \centering Digital Media Lab \\ University of Bremen \\ Bremen, Germany}
\and
Donald Degraen\thanks{e-mail: donald.degraen@canterbury.ac.nz}\\ %
\parbox{1.5in}{\scriptsize \centering HIT Lab NZ \\ University of Canterbury \\ Christchurch, New Zealand}
}

\teaser{
  \centering
  \includegraphics[width=\linewidth]{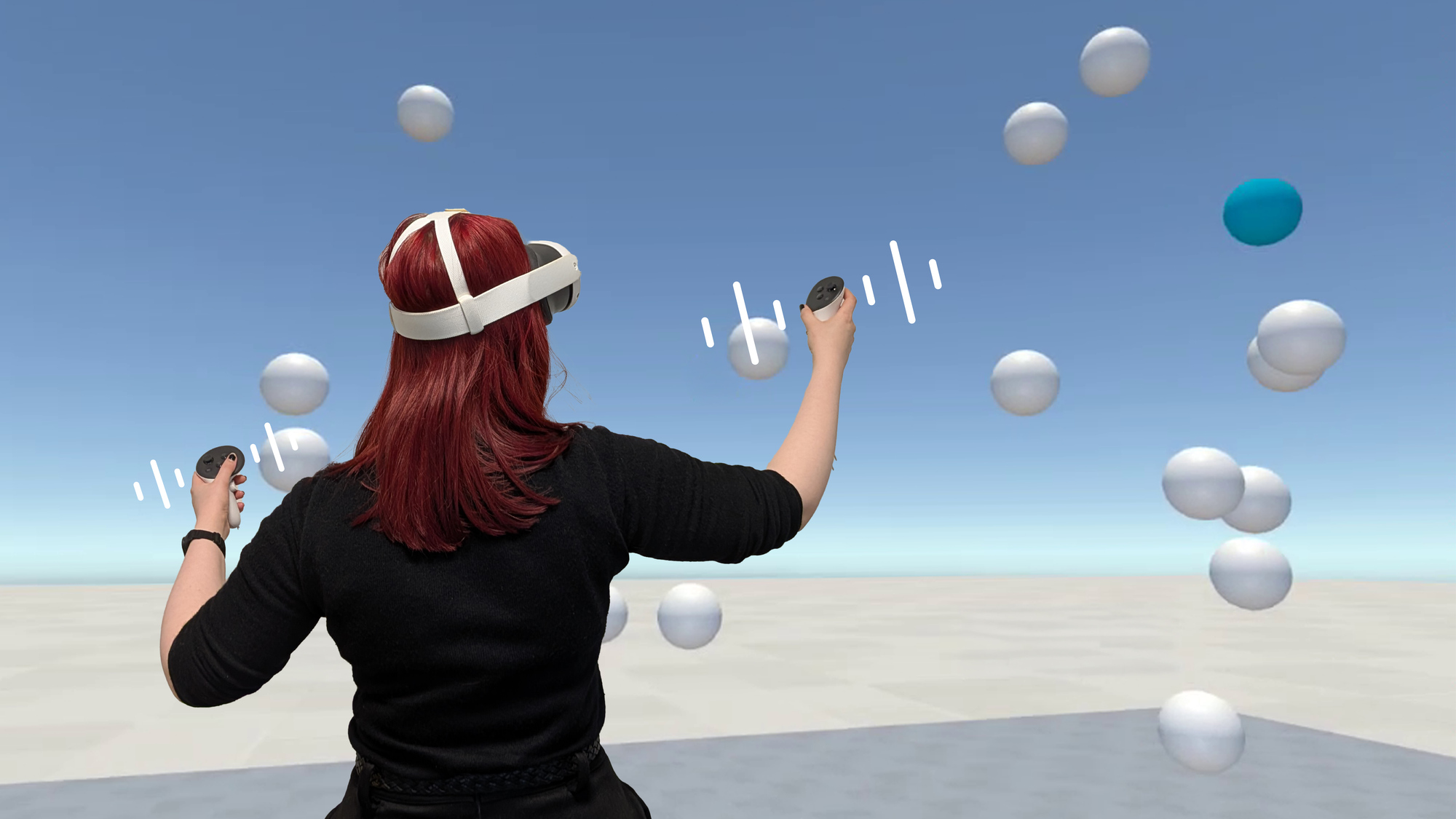}
  \caption{A user wearing a VR headset holds two controllers that deliver adaptive vibrotactile feedback, guiding them to locate a moving target in 3D space. Vibration intensity and frequency encode the object’s relative position, extending auditory-like localization cues into the tactile domain.}
  \label{fig:teaser}
}

\abstract{
Visual search is crucial in daily life, from scanning for relevant information to spotting signs of danger. When sensory channels are overloaded or degraded, cognitive tasks can be supported by crossmodal information representations through vibrotactile cues. We introduce Tactile Search, an approach that uses modulation of frequency and amplitude of vibrations to the hands, for guiding attention to the location of objects in 3D space. We evaluated this approach in a competitive VR game where participants searched for targets using both vision and touch. Across two studies -- an in-the-wild demonstration (n=55) and a controlled laboratory experiment (n=28) -- we found that vibrotactile feedback significantly improved performance and increased user confidence. In the combined haptic condition, performance did not differ across target heights. We further analyzed participants' subjective experiences and search strategies highlighting the benefits of the tactile cues. Our findings suggest that Tactile Search can enhance interaction and provide design considerations for integrating haptic search into interactive systems.
 }

\keywords{Haptic, Tactile, Sensory substitution, Spatial perception, Localization}

\begin{document}


\firstsection{Introduction}

\maketitle

Accurately targeting objects in three dimensional space is essential for interaction in everyday environments, supporting tasks such as orienting to approaching hazards, locating items outside the field of view, and maintaining spatial awareness when attention is divided across multiple stimuli. These abilities rely primarily on vision and audition. However, performance often declines when sensory resources are overloaded, divided, or degraded, a pattern predicted by perceptual load and multiple resource theories~\cite{wickens2002multiple, lavie1995perceptual, lavie2014blinded}. 
Empirical studies confirm similar effects in applied contexts such as navigation and driving, where sensory overload reduces situational awareness and increases the likelihood of missed events~\cite{chhan_-vehicle_2019, krasovsky2024understanding}. Such limitations motivate the need for alternative or complementary sensory channels to maintain efficient spatial search.

Haptic feedback has been explored as a supplemental channel for spatial guidance and has reduced search or response time in visually demanding tasks~\cite{lehtinen_dynamic_2012, tivadar_digital_2022, tan_haptic_2009, lindeman12003effective}. 
Vibrotactile cues in VR have supported invisible-object search and spatial sound localization~\cite{Beese2025Feel, Chelladurai2024SoundHapticVR}. More broadly, vibrotactile systems have been used to present directions and spatial warnings in navigation and vehicle interfaces~\cite{van_erp_presenting_2005, ho2005using, gaffary_use_2018}. These findings demonstrate that haptics as a modality to communicate spatial information can complement vision and audition during complex search tasks.

Haptic spatial guidance in VR/AR/XR can allow for conveying location information without adding to the visual load ~\cite{Marquardt2020Comparing,Trepkowski2022Multisensory}. Furthermore, the use of the already available controllers, offers a way to achieve this without requiring additional hardware. Haptic spatial guidance also serves accessibility: spatial audio may be inaccessible or insufficient for deaf and hard-of-hearing users~\cite{Chelladurai2024SoundHapticVR}, while haptic cues can provide blind users with non-visual access to spatial information~\cite{Li2024Comparing}. Safety-critical XR applications, such as surgical AR, may similarly benefit from guidance that does not compete for visual attention~\cite{Zhang2022Towards}.

Recent work modulated the amplitude of vibrotactile feedback delivered through handheld controllers according to their relative position to a target, enabling users to accurately localize a stationary object on the horizontal plane~\cite{wald_spatial_2025}.
Building on principles of sensory substitution~\cite{amedi_tactile_2023, maimon_case_2022, teng_seeing_2025}, this method employed relative vibration strength as an analog to interaural level differences in auditory localization, allowing users to infer the direction of an object also outside the field of view~\cite{wald_spatial_2025}. 

Expanding upon the previous research, this work evaluates visual search performance, and considers a dynamic environment with moving objects. Furthermore, the present work extends the prior horizontal amplitude method along two axes: from static localization to dynamic search, and from horizontal-only to including vertical via frequency modulation. In this context, it is important to clarify that in cognitive psychology, dynamic search is considered to be a different task from static localization~\cite{bennett_assessing_2021, Crowe2021Motion, Pylyshyn1988Tracking}. 

The task is a controlled abstraction of real-world spatial search and guidance demands in dynamic environments. Related demands arise in common use cases such as in-vehicle hazard cueing~\cite{ho2005using}, human-swarm supervision~\cite{Celebi2025Brief}, and teleoperation and collision avoidance~\cite{Huang2024AeroHaptix}. While other tasks are conceivable, our goal was to evaluate whether controller-based tactile cues can support search by shifting some spatial guidance away from vision and toward touch.

While our implementation also included the novel feature of height representation, as a first evaluation of visual, dynamic search, we measured the system against a vision-only baseline in accordance with similar visual-search studies~\cite{Beese2025Feel,lehtinen_dynamic_2012}.

In this work, we investigate how vibrotactile feedback delivered through commodity handheld devices can support targeting in dynamic search tasks. We examine how auditory-localization-inspired haptic feedback influences search performance relative to vision alone, how such cues shape users' strategies, confidence, and decision making, and whether frequency modulation can convey vertical spatial information.

This work is guided by the following research questions:

\begin{enumerate}
\item RQ1: To what extent does spatial vibrotactile feedback influence performance and subjective confidence during dynamic visual search tasks compared to vision alone?
\item RQ2: Does frequency based vibrotactile feedback affect users performance when locating targets at different vertical positions within three dimensional space?
\item RQ3: Do spatial vibrotactile cues shape search behavior, including turning direction, switching movements, and overshooting during target acquisition?
\end{enumerate}

To address these research questions, we implemented a spatial vibrotactile guidance method using modulated amplitude and frequency through commodity handheld controllers, applied in a competitive visual search game. We evaluated the method through two complementary studies, combining public demonstration with controlled laboratory analysis. Across these studies, we examined both performance and behavioral responses to understand how users interpret and act upon spatial haptic signals during search. Based on this investigation, this work contributes:

\begin{itemize}
\item A spatial vibrotactile guidance approach for dynamic 3D search, associated with faster targeting and higher confidence across a public study (n=55) and a controlled lab experiment (n=28); 
%\item An empirical extension of Spatial Haptics to active three dimensional search tasks, demonstrating how spatial vibrotactile cues influence targeting performance and subjective confidence relative to vision only conditions.
\item A first evaluation of frequency-based vibrotactile cues for representing vertical position during search, showing that a combined haptic condition evened out a height-dependent performance differences seen without haptic guidance. 
\item A behavioral analysis showing that this approach reduces unnecessary switching, overshooting, and improves initial turn direction, identifying strategy changes that can explain observed speed gains.
\end{itemize}

\section{Related Work}
Our approach builds on three main areas of prior work. The first concerns sensory load and multimodal search, showing how capacity limits restrict performance in visually demanding environments. The second addresses haptic feedback as a channel for directing attention and guiding search when other modalities are busy or unreliable, and the third involves sensory substitution.

\subsection{Sensory Overload and Multimodal Search}

Research on spatial perception and multimodal search establishes the need for alternative sensory channels when vision and audition become overloaded or unreliable. Work in perception and attention shows that spatial-task performance can decline when concurrent tasks compete for shared processing resources, while perceptual load changes the processing of irrelevant stimuli~\cite{wickens2002multiple, lavie1995perceptual, lavie2014blinded}. These constraints are evident in applied domains: reading through smart glasses while walking reduces gait and reading performance~\cite{krasovsky2024understanding}, while tactile in-vehicle displays have been proposed as a way to offload visual information and support situation awareness~\cite{chhan_-vehicle_2019}. When display complexity is high, distributing information across modalities can improve efficiency by shifting selected information to less burdened channels.

\subsection{Haptic Feedback for Spatial Guidance and Search}
Haptic feedback has been widely explored as a complementary channel for guiding attention and supporting spatial tasks when vision or audition are constrained. Vibrotactile systems have been used for orientation and navigation, reducing reliance on continuous visual monitoring~\cite{van_erp_presenting_2005, ho2005using, gaffary_use_2018}. In visually demanding conditions, directional haptic cues can improve spatial awareness and search speed by indicating where to look next rather than what to look for~\cite{lehtinen_dynamic_2012, tan_haptic_2009, lindeman12003effective}. Across localization and search tasks, prior work has shown that vibrotactile cues can represent direction, distance, and elevation through variations in stimulation location, amplitude, frequency, and temporal patterning~\cite{Gongora2017Experiments,Cabaret2022Perception,Nonino2021Subtle,JesusOliveira2017Designing,Beese2025Feel,wald_spatial_2025}.
Several interfaces provide haptic guidance for locating targets or directing hand movement using wearable or contactless technologies. Prior work includes midair ultrasound haptics~\cite{vo2015touching, freeman2019haptiglow, zhu2025knuckleguide, suzuki2019midair} and vibrotactile patterns delivered through gloves or wrist devices for guiding hands toward targets~\cite{trant_enhancing_2025, tajdari_navigating_2025, gunther_tactileglove_2018}. These systems demonstrate that haptic cues can effectively encode direction, but many rely on specialized hardware and are typically evaluated in constrained guidance tasks.  

\subsection{Sensory Substitution and Crossmodal Spatial Perception}
Sensory substitution research has shown that spatial information acquired through one modality can be remapped to another and used for navigation and object localization. Classic work demonstrated that tactile stimulation on the body can encode visual spatial layouts~\cite{bach-y-rita_vision_1969}. More recent work has shown that auditory spatial cues can be mapped onto haptic stimulation devices to improve sound localization~\cite{fletcher_electro-haptic_2020, fletcher_haptic_2020, fletcher_sensitivity_2021}. Devices such as the Tactile Vision Substitution System and tongue based electrotactile displays illustrate how users adapt to new sensorimotor mappings when feedback preserves underlying spatial structure~\cite{bach-y-rita_vision_1969, kaczmarek_tongue_2011, chebat_tactile-visual_2007}. Together, these findings support the feasibility of representing spatial location through vibrotactile patterns, provided that mappings are consistent and learnable. More recently, a sensory substitution-based method using amplitude differences between two handheld controllers demonstrated horizontal localization of static targets in VR~\cite{wald_spatial_2025}.

\section{Tactile Search}

Tactile Search uses amplitude modulation following the method proposed in \cite{wald_spatial_2025} and a novel frequency modulation to encode vertical position, enabling localization of objects across the full 3D space using two handheld controllers. We implemented and evaluated Tactile Search in a competitive VR search game designed to measure targeting performance and behavioral responses under naturalistic search conditions.

\subsection{Design and Implementation}

Our implementation is based on the hand-based localization method introduced by Wald et al.~\cite{wald_spatial_2025}.
Their algorithm simulates interaural level differences used in auditory localization by adjusting vibration amplitude based on the relative positions of the hands to a target object.
Within the user's personal arm reach, maximum amplitude is applied when a controller is nearest to the target, gradually decreasing to no-vibration at the furthest point from the target.
This scaling follows a non-linear curve so that differences are emphasized at closer ranges, similar to how auditory signals decrease in amplitude as they travel through space.
This provides an intuitive sensory substitution approach for representing spatial information through touch, which can easily be extended with complex signals for varying application use-cases.

While the existing implementation implicitly encodes altitude between the user's hands and the target object, the modulation of amplitude caused by height differences remains subtle in nature.
Therefore, we chose to extend this representation by associating height with a change in frequency.
Higher frequencies indicate elevated positions, whereas lower frequencies correspond to lower ones.
This relies on a well-known auditory crossmodal correspondence that associates higher pitches with higher spatial locations~\cite{chiou2012cross}.
It was chosen as an appropriate method for a system that simulates auditory localization, as well as marked as an opportunity to further study these correspondences in tactile in future studies.

Our implementation uses three fixed heights, using three fixed frequency signals.
A baseline vibration of 125 Hz was selected for horizontal cues at eye level.
This frequency was selected as it corresponds to peak human sensitivity for grip-based vibration~\cite{griffin2012frequency}.
Vertical positioning was represented by adjusting frequency from this baseline with a 90Hz signal for low targets and 150Hz for high targets.
The implementation was optimized for targets at a fixed distance of three meters, ensuring consistent intensity scaling and smooth real-time operation.

\subsection{Game Implementation}
\label{sec:game-implementation}

\begin{figure*}[t!]
    \centering
    \includegraphics[width=1\linewidth]{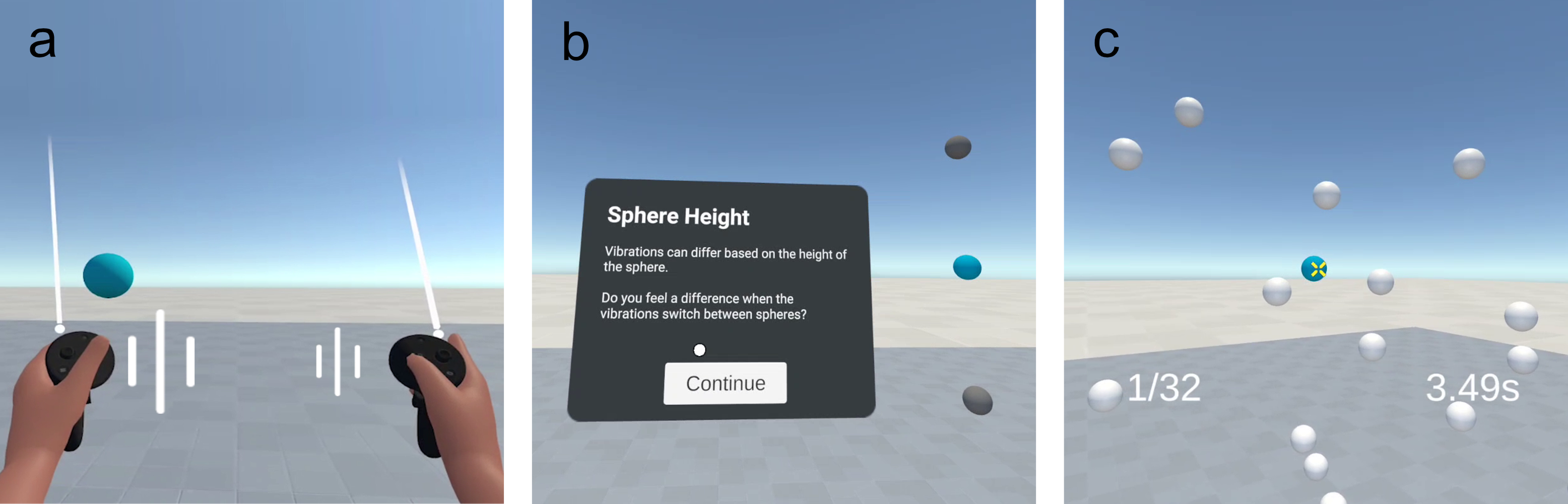}
    \caption{Visual Search Game. Users are (a) introduced to the concept of spatial vibrations through a moving target, and (b) frequency differences indicating different height levels. Afterwards, (c) they are tasked to find and select the blue sphere in several trials with and without spatial vibrations.}
    \label{fig:procedure}
\end{figure*}

To evaluate localization performance, we implemented our algorithm in a competitive visual search game, see \autoref{fig:procedure}.
The game was developed in Unity3D (version 2022.3.32f1) and deployed on a Meta Quest 3 headset.
The Meta Quest 3 controllers use wide-band voice coil actuators that enable modulation of both amplitude and frequency separately. The haptic condition combined vertical and horizontal cues.

The game starts with a welcome screen where users enter their details and are introduced to the concept.
A guided introduction familiarizes them with the method: a single vibrating sphere is used to show how vibration changes with hand movement, three spheres at different heights demonstrate how frequency encodes elevation, a circling sphere illustrates how vibration varies with movement around the participant, and finally, two spheres are presented where only one vibrates, requiring users to select it by aiming a crosshair at the target and pressing any controller button.
To increase accessibility, the introduction provides a voice-over of all text displayed throughout.

After the introduction, users are positioned at the center of a scene where $80$ randomly distributed gray spheres continuously circle the user.
Spheres are positioned at a random vertical offset of $-1.2$ and $+1.2$ meters calculated from the user's eye level, and at a random radius between $2.8$ and $3.3$ meters from the center.
They are randomly assigned a clockwise or counterclockwise rotation around the center with a random speed between $10^\circ$ and $20^\circ$ per second.

After an initial countdown, the game starts.
For each trial, the user is tasked to find and select a blue colored sphere as quickly as possible.
To make a selection, they align the yellow crosshair at the center of their view with the sphere and press any controller button as they did at the end of the introduction stage.
When a target is chosen, it disappears and the next trial begins immediately.
The user's view shows the trial number and a timer that resets for each trial.

For each trial, the target blue sphere appears outside the user's field of view, with its horizontal position determined by selecting a random angle behind the user between $135^\circ$ and $225^\circ$ from the user's gaze at $0^\circ$.
The target is positioned at a radius of $3.0$ meters from the center and at one of three distinct heights, i.e., low, middle, or high.
The middle height is defined by the user's eye height, while the low and high heights offset this vertical position by $-1.0$ and $+1.0$ meters respectively.
Similar to the other spheres, the target is randomly assigned a clockwise or counterclockwise rotation around the center with a random speed between $10^\circ$ and $20^\circ$ per second.

Across the session, a total of $32$ targets are presented to the user.
The first two targets are considered warm-up targets and are excluded from further analysis.
The first one is presented with vibrations and the second without.
This was implemented to allow users to familiarize themselves with the task and make sure they do not wait for a haptic signal.
The following $30$ targets are randomly assigned to either haptics condition such that $15$ targets have vibrations and $15$ do not.
For each condition, $5$ targets are presented at the medium height, $5$ at the low height, and $5$ at the high height.

After each session, users are shown a comparison of their performance in both conditions, highlighting the difference between trials where vibrations supported localization and those where no haptic guidance was present.

\section{Experiment I: Public Demonstration}

To evaluate our approach, we showcased our visual search game through a public demonstration at a major conference on Human-Computer Interaction (HCI).
Specifically, we aimed to understand if the addition of spatialized vibrotactile signals increased visual search performance, answering \textbf{RQ1}.

\subsection{Participants}

During our public interactive demonstration, we invited all passers-by to take part in our interactive game.
Upon starting the game, participants were asked if their data could be used for further analysis.
A total of 55 consenting participants completed our game from start to end.
Due to the nature of the study, we were not able to collect demographic information.

\subsection{Ethics} 
For our study design, we followed the ethical guidelines of our institution and received approval from the ethics committee of the University of Duisburg-Essen. 
All participants in our studies were informed about the procedure and the data protection policy and provided informed consent.

\subsection{Experimental Design}

Our study consisted of a within-subjects design where all participants selected target objects in a competitive visual search game.
As independent variables, we distinguish the presence of vibrotactile cues (haptics vs. no haptics), and the height of a target (low vs. medium vs. high).
As dependent variables, we identify the time required to find and select a target (duration).

\subsection{Procedure}

The demonstration consisted of a demo booth with further information on the concept and implementation.
Interested participants were first briefly introduced to the concept of tactile spatial localization through sensory substitution.
The experimenter explained that the experience would use vibrations in handheld controllers to mimic auditory localization.
After donning and adjusting the VR headset, participants followed the game procedure as explained in \autoref{sec:game-implementation}.
To incentivize participation, a separate public display showed a scoreboard with the top 10 user scores for trials with haptics.
Auditory cues from the controllers were not considered an affecting factor due to the masking sounds in the loud public setting of the experiment.

\subsection{Measurements}

During our public demo, we collected the following measurements:

\begin{itemize}
    \item Trial Duration: the time in seconds required to find and select a target sphere;
    \item Subjective Confidence: user's subjective confidence level of finding and selecting targets with or without haptics, measured on a 10-point scale (1 - strongly disagree, 10 - strongly agree).
\end{itemize}

%\textcolor{red}{Did we collect other measurements?}

% In the public demo, we assessed users' confidence using a 1–10 Likert scale to capture fine-grained variations in self-reported certainty. Speed was quantified by the time taken to successfully locate each target, providing an objective performance measure.

\subsection{Results}

The results of our public demonstrator show that haptic feedback significantly improved performance across conditions while also increasing participants' subjective confidence.
Height differences influenced duration in a quadratic fashion, indicated by the medium level being significantly faster than the high and low.
% No evidence was found for an interaction between haptics and height.

\begin{figure*}[t!]
    \centering
    \begin{subfigure}[t]{0.18\textwidth}
        \centering
        \includegraphics[width=1\linewidth]{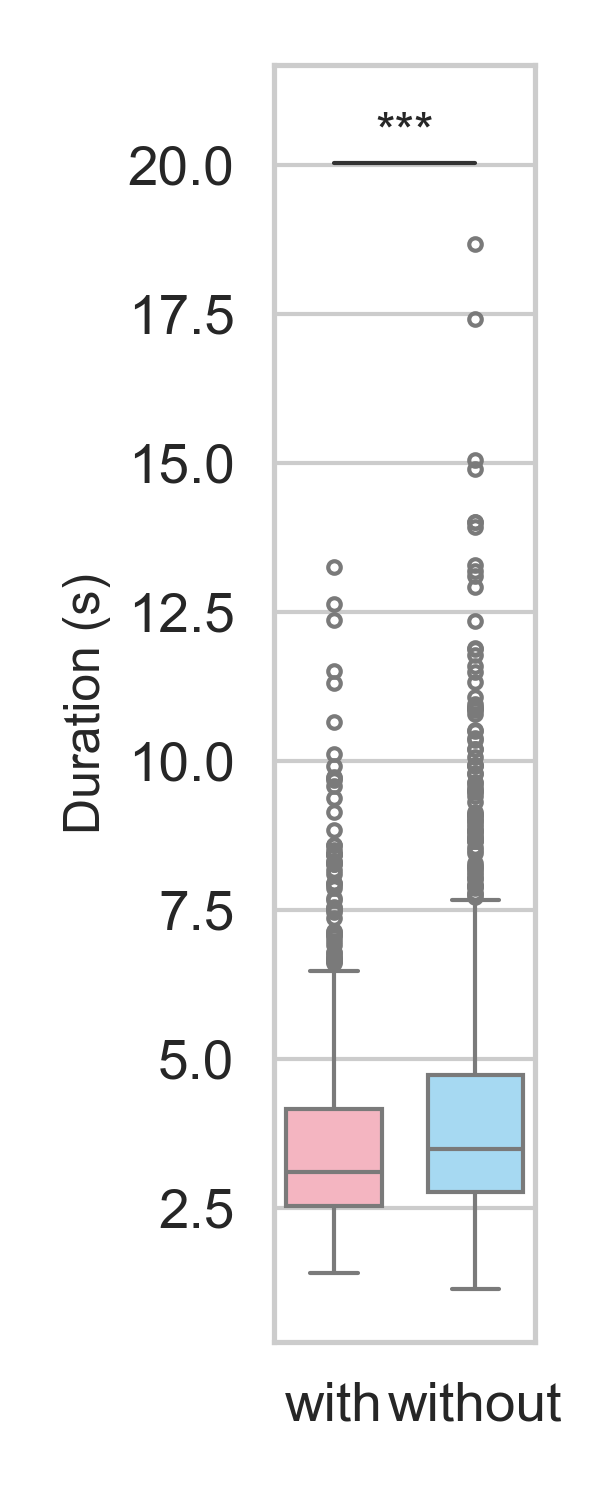}
        \caption{Haptics}
    \end{subfigure}%
    ~ 
    \begin{subfigure}[t]{0.27\textwidth}
        \centering
        \includegraphics[width=1\linewidth]{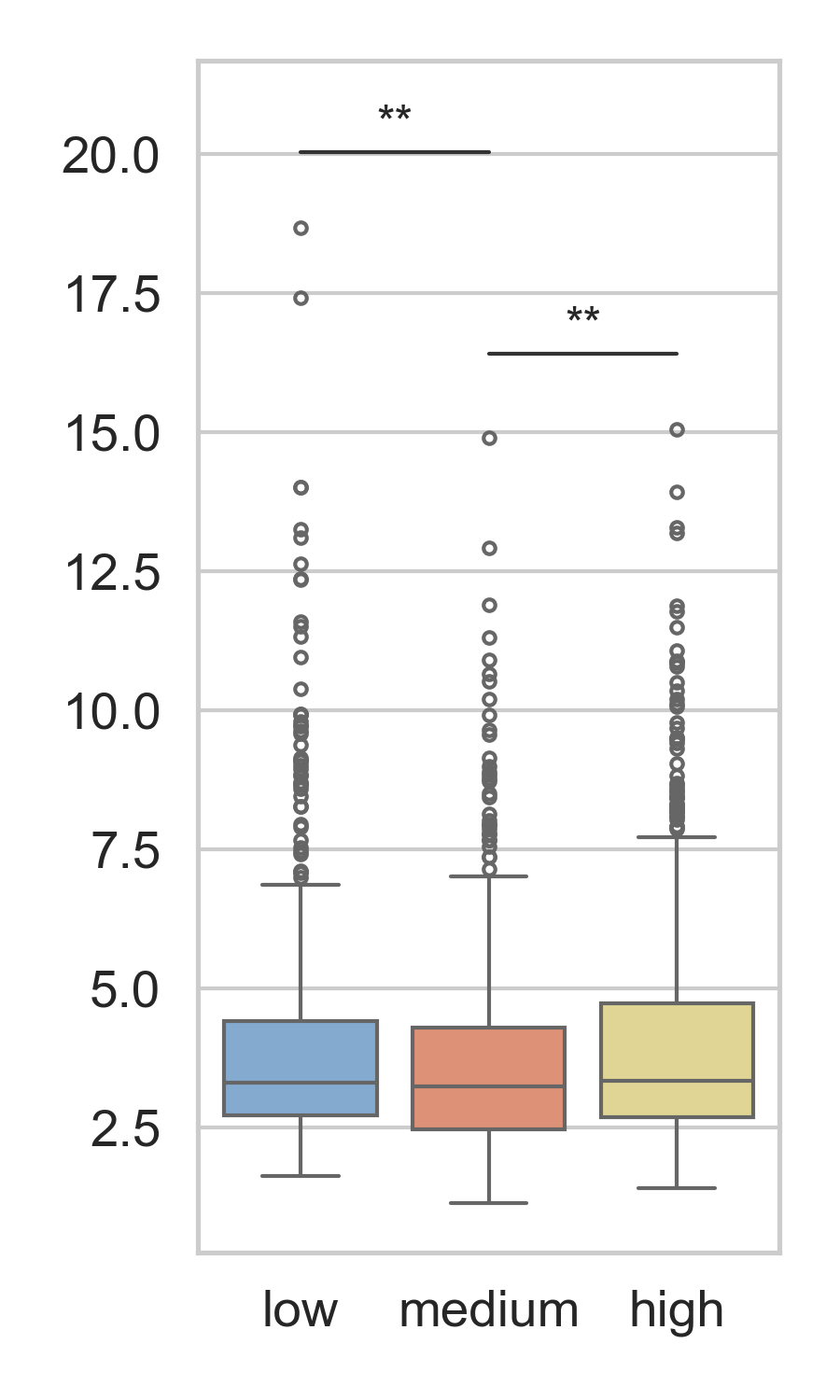}
        \caption{Target Height}
    \end{subfigure}%
     ~ 
    \begin{subfigure}[t]{0.18\textwidth}
        \centering
        \includegraphics[width=1\linewidth]{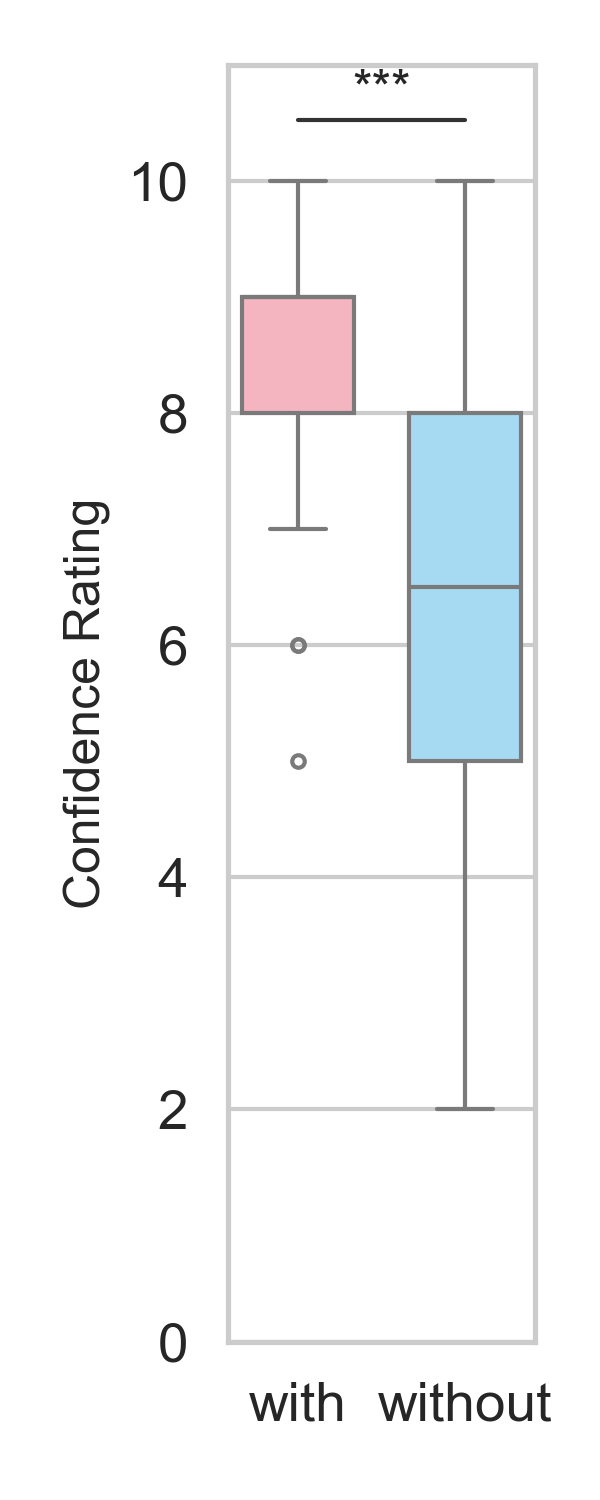}
        \caption{Confidence Ratings}
    \end{subfigure}%
    \caption{Boxplots depicting trial duration (s). Here, (a) contrasts the duration of selecting a target with and without haptic feedback, (b) contrasts the difference between different target heights and combines haptics and no-haptic data, and (c) subjective confidence ratings.}
    \label{fig:demo-results}
\end{figure*}

\begin{table}
    \centering
    \caption{Visual Search Performance.}
    \begin{tabular}{r|cc}
							& \multicolumn{2}{c}{\textbf{Search Time (s)}}	\\
						 	&  Haptics 				& No Haptics			\\
	\textbf{Target Height}	& $\bar{x}$ / $\sigma$	& $\bar{x}$ / $\sigma$	\\ \hline
	High					& 3.71 / 1.68			& 4.42 / 2.52			\\
	Medium					& 3.39 / 1.45			& 4.01 / 2.10			\\
	Low						& 3.77 / 1.93			& 4.27 / 2.51			\\ \hdashline
	\textbf{Overall}		& 3.62 / 1.70			& 4.23 / 2.39			\\
    \end{tabular}
    \label{tab:demo-metrics}
\end{table}

\subsubsection{Confidence Ratings}
Participants rated how confident they felt in performing the task in each condition.
For targets where vibrations were active, subjective confidence levels were on average $8.58$ (Median = 9, SD = $1.15$, Min = $5$, Max = $10$).
For targets without vibrations, subjective confidence levels were on average $6.40$ (Median = 6, SD = $2.04$, Min = $2$, Max = $10$) (see  \autoref{fig:demo-results}).

Wilcoxon signed-ranks tests comparing subjective confidence levels between haptics vs. no haptics conditions indicated a significant difference (${Z = -4.968}$, ${p < .001}$).
These results show that participants were significantly more confident in finding the target when vibrotactile cues were active.

\subsubsection{Performance Ratings}
To evaluate performance, we assessed the time required to select a target as trial duration, see \autoref{tab:demo-metrics} and \autoref{fig:demo-results}.
A Shapiro-Wilk test on the residuals of the dependent variable (duration) indicated a statistically significant difference (${p < .001}$, ${W = .801}$), meaning that the dataset does not show a normal distribution.
In order to improve normality and homogeneity of variance, we applied a Box–Cox transformation on the dependent variable, with an estimated $\lambda$ of $-0.6308$.
All subsequent analyses were conducted on the transformed data.
A further Shapiro-Wilk test on the residuals of the transformed values indicated a statistically significant difference (${p < .001}$, ${W = .996}$).
However, the distribution of residuals showed a small skewness ($-.10$) and kurtosis ($-.160$), as confirmed by visual inspection of the histogram and QQ-plot.
Repeated-measures ANOVA is generally robust to modest non-normality, although robustness depends on the distribution and design~\cite{blanca2023nonnormal}.
Therefore, we conducted a two-way repeated-measures ANOVA with \textit{haptics} (haptics vs.\ no haptics) and \textit{height} (high, medium, low) as within-subjects factors on duration.

Results indicated a significant main effect of \textit{haptics} ($F(1, 54) = 34.47$, $p < .001$, $\eta_p^2 = .390$).
Trials with haptics took on average $3.62$ seconds ($SD = 1.70$), while trials without haptics took on average $4.23$ seconds ($SD = 2.39$).
These results show that trials with haptic feedback were significantly faster than those without.

A significant main effect of \textit{height} was also observed ($F(2, 108) = 8.70$, $p < .001$, $\eta_p^2 = .139$). 
Pairwise comparisons with Bonferroni adjustment showed that the medium height condition ($M = 3.70$, $SD = 1.83$) was significantly faster than both high ($M = 4.07$, $SD = 2.17$; $p = .004$) and low ($M = 4.02$, $SD = 2.25$; $p = .002$).
No difference was found between high and low height conditions ($p = 1.000$). A value of $p = 1.000$ reflects a Bonferroni-adjusted pairwise comparison, where adjusted values above 1 are capped at 1.000 by the statistical software. Bonferroni corrections were applied to the follow-up pairwise comparisons.
Planned contrasts revealed that the effect of height was quadratic ($F(1, 54) = 14.78$, $p < .001$, $\eta_p^2 = .215$) reflecting the faster search time in the medium condition relative to high and low.

% The interaction between \textit{haptics} and \textit{height} was not significant ($F(2, 108) = 0.25$, $p = .778$, $\eta_p^2 = .005$).
% Thus, the effect of haptic feedback on task duration did not differ across heights.

\subsection{Summary}
Haptic feedback significantly improved overall performance, reducing task duration compared to trials without haptics.
Additionally, users felt subjectively more confident in finding and selecting targets in the presence of haptic cues.
Differences in target height significantly influenced task duration, showing that targets at eye level were easier to find and select than targets presented at higher or lower locations.

\section{Experiment II: Lab Study}

% In a subsequent lab study, we \textcolor{green}{aimed to understand if frequency modulation of spatialized vibrotactile signals increases vertical visual search performance, answering \textbf{RQ2}}, and how such haptics improve visual search behavior, addressing \textbf{RQ3}.
In a subsequent lab study, we aimed to understand if spatialized vibrotactile signals increase vertical visual search performance, answering \textbf{RQ2}, and how such haptics improve visual search behavior, addressing \textbf{RQ3}.

\subsection{Participants}

We recruited a separate cohort of $28$ participants (7 female, 21 male, 0 other) for our lab study through on-campus outreach.
Participants were aged between 21 and 39 years old ($M = 26$ years, $SD = 4.5$). $21$ participants indicated that they are right-handed, while 5 were left-handed and 2 were ambidextrous. With respect to VR use, 10 participants reported they use VR rarely, 5 occasionally, 6 frequently, and 7 very often.
Informed consent was provided in writing through a pre-study survey.

Based on a power analysis using the haptics effect size from Study 1 ($\sim 0.8$), $n=28$ provides high power ($\sim .98$) to detect a comparable haptics effect at $\alpha=.05$.

\subsection{Ethics} 
For our study design, we followed the ethical guidelines of our institution and received approval from the ethics committee of the University of Duisburg-Essen. 
All participants in our studies were informed about the procedure and the data protection policy and provided informed consent.

\subsection{Experimental Design}

The study consisted of a within-subjects design where all participants selected target objects in a competitive visual search game.
As independent variables, we distinguish the presence of vibrotactile cues (haptics vs. no haptics), and the height of a target (low vs. medium vs. high).
As dependent variables, we identify the time required to find and select a target (duration).

\subsection{Procedure}

Before the experiment, consent forms were prepared, and participant IDs assigned. 
The protocol was explained to the participants, and they were told to take their time during the onboarding to get to know the system. They were instructed that during the game they should locate both vibrating and non vibrating targets as fast as possible. Participants adjusted the VR headset and then recalibrated the view by facing forward and holding down the controller button.

After completing the VR phase, participants filled out a post-experiment questionnaire including Likert scales for confidence (1–10) and parameters of the haptic feedback (1–7), as well as short-answer prompts for qualitative feedback.

As this experiment was conducted in a controlled environment, there is a slight concern regarding the effects of auditory cues from the controllers. Though not present in formal documentation, to our knowledge the devices produce a very subtle buzz roughly in the range of 25–35 dBA, such that at arm’s distance even in a controlled environment the sound of the vibration is barely audible, and could not provide a reliable cue. This was also validated by the researchers personal experience.

\subsection{Measurements}

We collected the following measurements:

\begin{itemize}
    \item Trial Duration: the time in milliseconds required to find and select a target sphere;
    \item User Tracking: we logged user's head and controller movements at a rate of 8 Hz;
    \item Trial Tracking: we logged all in-game events and sphere locations at a rate of 8 Hz;
    \item Subjective Confidence: user's subjective confidence level of finding and selecting targets with or without haptics, measured on a 10-point scale (1 - strongly disagree, 10 - strongly agree);
    \item Subjective Haptics Contribution: user's subjective evaluation of how much the haptic feedback contributed to their experiences, measured on a 7-point scale (1 - strongly disagree, 7 - strongly agree);
    \item Qualitative Feedback: we provided short-answer prompts allowing participants to elaborate on their experience.
\end{itemize}

%\textcolor{red}{Check above. Did we collect other measurements?}

% In the follow-up lab experiment, we retained the 1–10 Likert scale for confidence and used the same continuous timing metric for speed to maintain comparability with the demo. Additionally, we introduced a separate 1–7 Likert scale specifically to evaluate how much the haptic feedback contributed to participants' experiences, and split to horizontal and vertical. This scale provided a clear midpoint to differentiate between experiences where haptics helped or hindered relative to participants' expectations. The midpoint ensured participants were not compelled into polarized responses. Finally, we incorporated short-answer prompts allowing participants to elaborate qualitatively, enriching our interpretation of the quantitative data.

\section{Results}

The results of our lab study show that haptic feedback significantly improved performance across conditions while also increasing participants' subjective confidence. Overall, no performance differences were found in different target heights. Interaction effects between haptics and target height showed performance differences across target heights in the no-haptics condition were not observed in the combined haptic condition.
Haptics also affected specific measures of visual search behavior, including reduction in directional switches and overshoots, and was associated with a higher probability of correct initial turns towards the target.
We furthermore outline qualitative feedback provided by participants detailing on their impressions.

\subsection{Confidence Ratings}

Participants rated how confident they felt in performing the task in each condition.
For targets where vibrations were active, subjective confidence levels were on average $8.18$ (Median = $8$, SD = $1.09$, Min = $6$, Max = $10$).
For targets without vibrations, subjective confidence levels were on average $5.25$ (Median = $5$, SD = $1.94$, Min = $1$, Max = $8$).

Wilcoxon signed-ranks tests comparing subjective confidence levels between haptics vs. no haptics conditions indicated a significant difference (${Z = -4.343}$, ${p < .001}$).
These results reinforce the results obtained from the public demonstrator and show that participants were significantly more confident in finding the target when vibrotactile cues were active.

\subsection{Performance Ratings}
\begin{figure*}[t!]
    \centering
    \begin{subfigure}[t]{0.18\textwidth}
        \centering
        \includegraphics[width=1\linewidth]{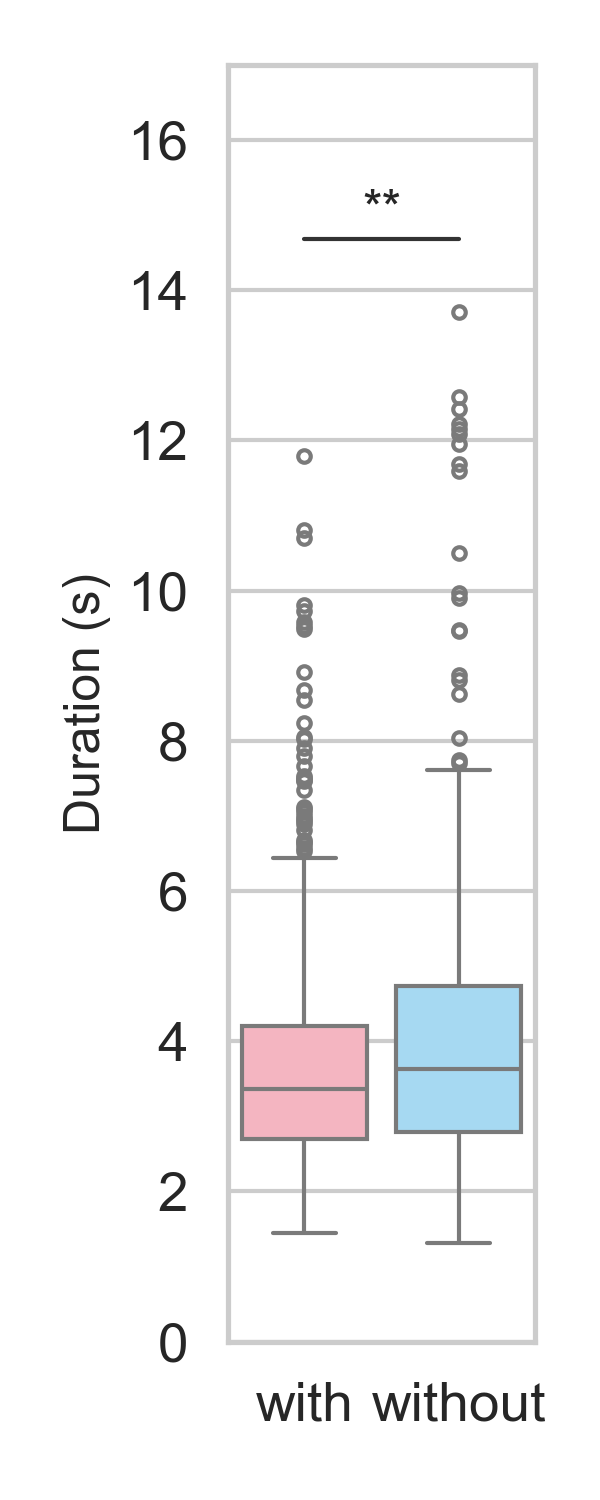}
        \caption{Haptics}
    \end{subfigure}%
    ~ 
    \begin{subfigure}[t]{0.27\textwidth}
        \centering
        \includegraphics[width=1\linewidth]{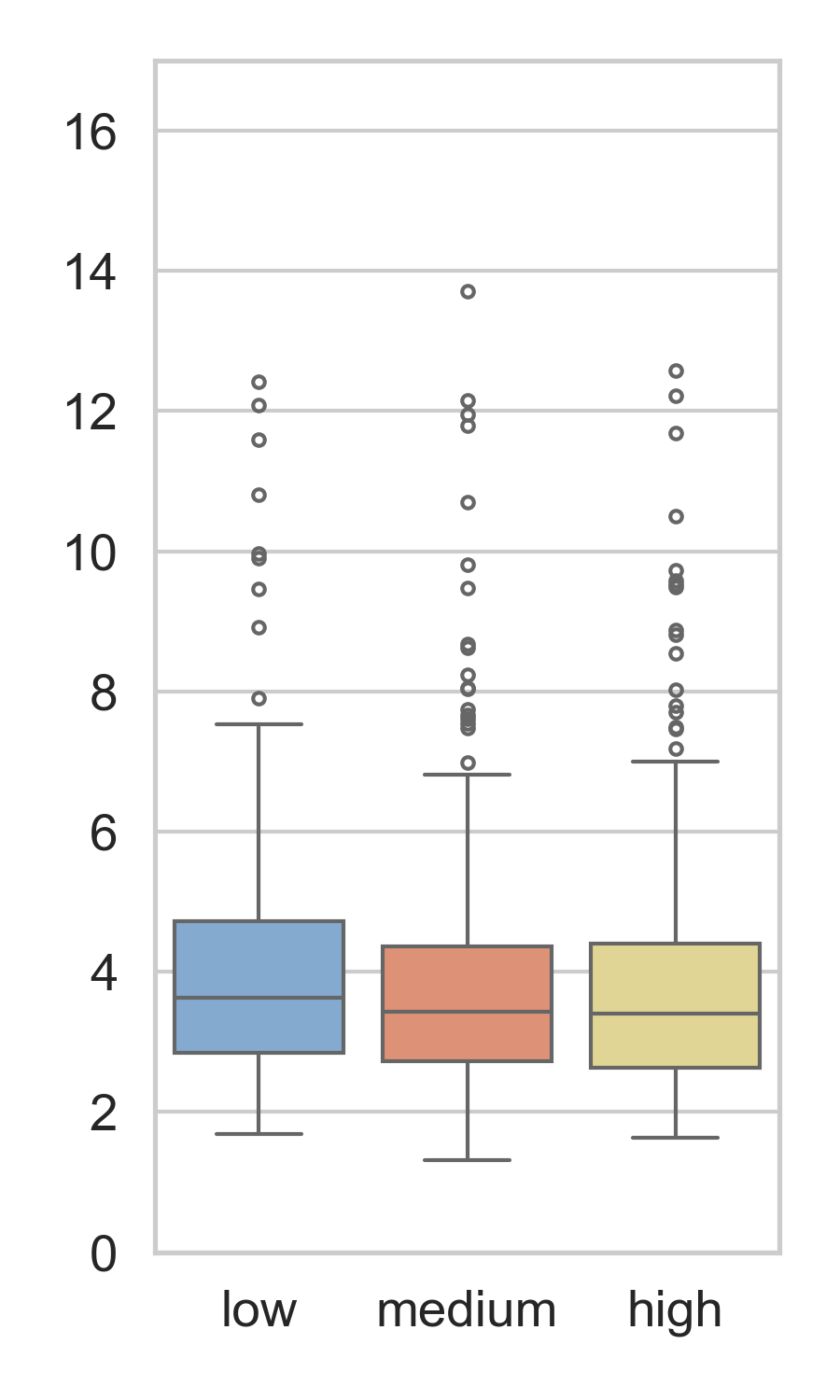}
        \caption{Target Height}
    \end{subfigure}%
    ~ 
    \begin{subfigure}[t]{0.54\textwidth}
        \centering
        \includegraphics[width=1\linewidth]{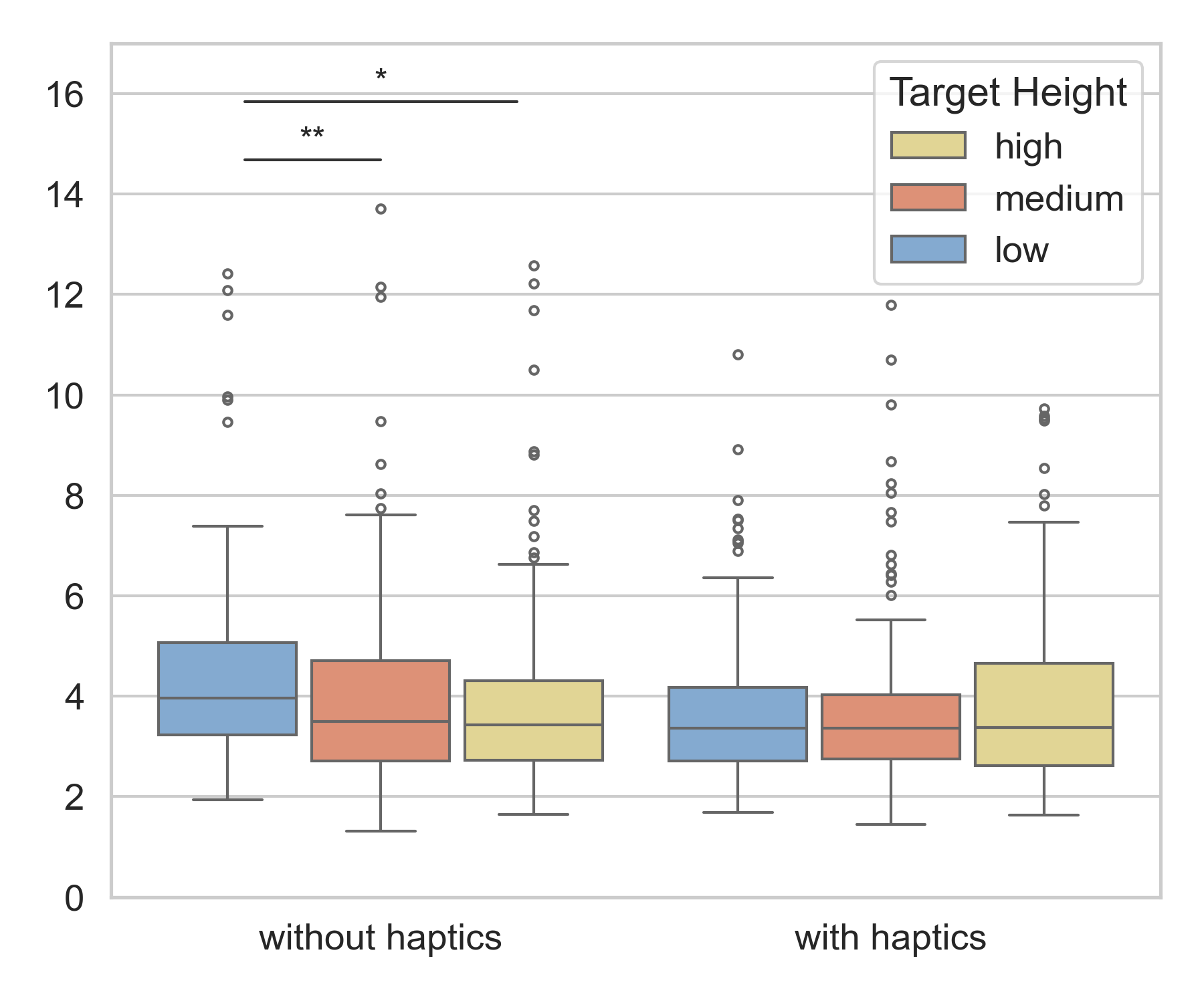}
        \caption{Target Height per Haptics Condition}
    \end{subfigure}
    \caption{Boxplots depicting Visual Search Performance. Here, (a) contrasts the duration of selecting a target with and without haptic feedback, (b) the difference between different target heights, and (c) the duration per target height for each condition.}
    \label{fig:study2-results}
\end{figure*}

To evaluate performance, we assessed the time required to select a target as trial duration, see \autoref{tab:study2-metrics} and \autoref{fig:study2-results}.
A Shapiro-Wilk test on the residuals of the dependent variable (duration) indicated a statistically significant difference (${p < .001}$, ${W = .830}$), meaning that the dataset does not show a normal distribution.
We transformed the dependent variable (duration) using a Box--Cox transformation to improve normality and homogeneity of variance.
The estimated transformation parameter was $\lambda = -0.5138$, and all reported results are based on the transformed data.
A further Shapiro-Wilk test on the residuals of the transformed values showed no statistically significant difference (${p = .365}$, ${W = .998}$), indicating a normally distributed dataset.
For analysis, we conducted a two-way repeated-measures ANOVA with \textit{haptics} (haptics vs.\ no haptics) and \textit{height} (high, medium, low) as within-subjects factors on duration, and consider both main and interaction effects.

\subsubsection{Main Effects}
Results indicated that trials with haptics took on average $3.79$ seconds ($SD = 1.67$), while trials without haptics took on average $4.11$ seconds ($SD = 1.95$).
We observed a significant main effect of \textit{haptics}, indicating that performance improved with haptic feedback compared to no haptics ($F(1, 27) = 4.94$, $p = .035$, $\eta_p^2 = .155$). Observed power was approximately .60 across all effects.

Overall, trials with targets at the medium height took on average $3.89$ seconds ($SD = 1.87$), at the high height took on average $3.91$ seconds ($SD = 1.87$) and low on average $4.05$ ($SD = 1.72$).
The main effect of \textit{height} was not significant ($F(2, 54) = 3.05$, $p = .056$, $\eta_p^2 = .101$), indicating that overall performance did not change depending on target height.

\subsubsection{Interaction Effects}
Results revealed a significant \textit{haptics $\times$ height} interaction ($F(2, 26) = 3.48$, $p = .046$, $\eta_p^2 = .211$).
Multivariate contrasts indicated a significant linear-by-linear interaction ($F(1, 27) = 6.36$, $p = .018$, $\eta_p^2 = .191$).
Follow-up comparisons showed that at the lowest height, performance improved significantly with haptics ($p < .001$), whereas at high and medium heights, no differences between haptics and no haptics were found.
Additionally, within the no-haptics condition, the low height was significantly slower than medium and high ($p = .002$ and $p = .024$, respectively).

The interaction between \textit{haptics} and \textit{height} revealed a meaningful pattern. 
In the no-haptics condition, performance at the low height was significantly worse than at both medium and high. 
By contrast, in the haptics condition, no differences between height levels were observed (all $p = 1.000$). A value of $p = 1.000$ reflects a Bonferroni-adjusted pairwise comparison, where adjusted values above 1 are capped at 1.000 by the statistical software. Bonferroni corrections were applied to the follow-up pairwise comparisons. 
In the combined haptic condition, no performance disadvantage was observed for the low height condition.

\begin{table*}
    \centering
    \caption{Visual Search Performance and Patterns.}
    \begin{tabular}{r|cc|cc|cc|cc}
							& \multicolumn{2}{c|}{\textbf{Search Time (s)}}		& \multicolumn{2}{c|}{\textbf{Horizontal Switches}}		& \multicolumn{2}{c|}{\textbf{Vertical Switches}}		& \multicolumn{2}{c}{\textbf{Overshoots}}				\\
						 	&  Haptics 				& No Haptics				&  Haptics 				& No Haptics					&  Haptics 				& No Haptics					&  Haptics 				& No Haptics					\\
	\textbf{Height}			& $\bar{x}$ / $\sigma$	& $\bar{x}$ / $\sigma$		& $\bar{x}$ / $\sigma$	& $\bar{x}$ / $\sigma$			& $\bar{x}$ / $\sigma$	& $\bar{x}$ / $\sigma$			& $\bar{x}$ / $\sigma$	& $\bar{x}$ / $\sigma$			\\ \hline
	High					& 3.91 / 1.80			& 3.91 / 1.94				& 18.6 / 12.4			& 18.6 / 14.6					& 17.8 / 11.6			& 17.3 / 13.1					& 3.34 / 2.16			& 2.96 / 1.98					\\
	Medium					& 3.74 / 1.71			& 4.05 / 2.01				& 19.7 / 12.9			& 18.4 / 10.5					& 17.8 / 14.5			& 18.8 / 12.9					& 2.76 / 1.95			& 2.75 / 1.80					\\
	Low						& 3.74 / 1.50			& 4.37 / 1.87				& 18.8 / 12.9			& 22.5 / 13.5					& 19.0 / 14.0			& 22.0 / 14.4					& 3.07 / 2.19			& 3.43 / 1.86					\\ \hdashline
	\textbf{Overall}		& 3.79 / 1.67			& 4.11 / 1.95				& 19.0 / 12.7			& 19.8 / 13.1					& 18.2 / 13.4			& 19.3 / 13.6					& 3.06 / 2.11			& 3.05 / 1.90					\\
    \end{tabular}
    \label{tab:study2-metrics}
\end{table*}

\subsection{Search Behavior}

We extracted all logged events and positions from the headset and evaluated directional switches in the horizontal and vertical planes, target overshoots, and initial turns.
For each metric, we applied a Box–Cox transformation on the dependent variable and report on the estimated $\lambda$.
All subsequent analyses were conducted on the transformed data.
Shapiro-Wilk test on the residuals of the transformed values were used to evaluate the normal distribution of the dataset.
For cases with statistically significant differences, we report skewness and kurtosis before continuing with repeated-measures ANOVAs. Bonferroni corrections were applied to follow-up pairwise comparisons. Adjusted values above 1 were capped at 1.000 by the statistical software.
The means and standard deviation are presented in \autoref{tab:study2-metrics}.

\subsubsection{Horizontal Switches}
During the study, we logged users' view angle at a rate of 8 Hz.
For each trial, we counted the amount of directional changes performed in the horizontal axis to determine the amount of horizontal switches.
The estimated transformation parameter of the Box-Cox transformation was $\lambda = 0.2772$.
The transformed data showed significant Shapiro-Wilk tests (${p < .001}$, ${W = .896}$), but small skewness ($.06$) and kurtosis ($2.83$).

In trials with haptics, participants performed an average of $19.0$ changes ($SD = 12.7$), and $19.8$ changes ($SD = 13.1$) in trials without haptics.
We did not find a significant main effect on haptics ($F(1, 27) = .465$, $p = .501$, $\eta_p^2 = .017$), indicating that the presence of haptics did not influence the amount of horizontal switches.

For targets presented at the high height, participants performed an average of $18.6$ changes ($SD = 13.5$), $19.0$ changes ($SD = 11.8$) for the medium height and $20.7$ changes ($SD = 13.3$) for the low height.
A significant main effect of \textit{height} was observed ($F(2, 54) = 3.23$, $p = .048$, $\eta_p^2 = .107$) indicating that the amount of horizontal switches changed depending on the height of the target object.
However, pair-wise comparisons did not reveal significant differences between different heights after correction.

Results revealed a significant \textit{haptics $\times$ height} interaction ($F(2, 54) = 5.06$, $p = .010$, $\eta_p^2 = .158$).
Follow-up comparisons revealed a meaningful pattern.
For targets presented at the lowest height, the amount of horizontal directional changes was significantly higher when no haptics were present than when haptics were present ($p = .010$).
For trials without haptics, low targets were related with a significantly higher amount of horizontal directional changes than the medium ($p = .024$) and the high targets ($p = .002$).
No differences were found for trials with haptics (all $p = 1.000$).

\subsubsection{Vertical Directional Changes}
During the study, we logged users' view angle at a rate of 8 Hz and counted the amount of directional changes performed in the vertical axis to determine the amount of vertical switches.
The estimated transformation parameter of the Box-Cox transformation was $\lambda = 0.1394$.
The transformed data did not show significant Shapiro-Wilk tests (${p = .191}$, ${W = .997}$), and small skewness ($-.007$) and kurtosis ($.316$).

In trials with haptics, participants performed an average of $18.2$ vertical changes ($SD = 13.4$), and $19.3$ changes ($SD = 13.6$) in trials without haptics.
We did not find a significant main effect on haptics ($F(1, 27) = 1.485$, $p = .234$, $\eta_p^2 = .052$), indicating that the presence of haptics did not influence the amount of vertical switches.

For targets presented at the high height, participants performed an average of $17.5$ changes ($SD = 12.4$), $18.3$ changes ($SD = 13.7$) for the medium height and $20.5$ changes ($SD = 14.2$) for the low height.
A significant main effect of \textit{height} was observed indicating that the amount of vertical switches changed depending on the height of the target object ($F(2, 54) = 5.86$, $p = .005$, $\eta_p^2 = .178$).
Pair-wise comparisons revealed a significant difference between the low and high targets ($p = .012$), but not between the low and medium ($p = .051$), and the medium and high ($p = 1.00$).

The interaction between \textit{haptics} and \textit{height} was not significant ($F(2,54) = 1.372$, $p = .262$, $\eta_p^2 = .048$)
Thus, the effect of haptic feedback on vertical directional changes did not differ across heights.

\subsubsection{Target Overshoots}
We defined the amount of overshoots as the amount of times the participant's center view passed by the target's position on the horizontal plane.
The estimated transformation parameter of the Box-Cox transformation was $\lambda = 0.2463$.
The transformed data showed significant Shapiro-Wilk tests (${p < .001}$, ${W = .900}$), but small skewness ($1.225$) and kurtosis ($2.491$).

In trials with haptics, participants overshot the target an average of $3.06$ times ($SD = 2.11$), and $3.05$ times ($SD = 1.90$) in trials without haptics.
We did not find a significant main effect on haptics ($F(1, 27) = .559$, $p = .461$, $\eta_p^2 = .020$), indicating that the presence of haptics did not influence the amount of overshoots.

For targets presented at the high height, participants overshot the target an average of $3.15$ times ($SD = 2.08$), $2.76$ times ($SD = 1.87$) for the medium height and $3.25$ times ($SD = 2.04$) for the low height.
A significant main effect of \textit{height} was observed indicating that the amount of overshoots changed depending on the height of the target object ($F(2, 54) = 3.27$, $p = .046$, $\eta_p^2 = .108$).
Pairwise comparisons indicated significantly more overshoots for high than medium targets ($p = .015$), but no difference between low and high targets ($p = 1.000$) or between low and medium targets ($p = .393$).

Results revealed a significant \textit{haptics $\times$ height} interaction ($F(2, 54) = 3.69$, $p = .031$, $\eta_p^2 = .120$).
Pair-wise comparisons indicated that for targets at the low height, trials with haptics were associated with a significantly lower amount of overshoots than trials without haptics ($p = .023$).
For trials without haptics, low targets were related with a significantly higher amount of overshoots than the medium ($p < .001$) and the high targets ($p = .046$).
No differences were found for trials with haptics (all $p = 1.000$, except medium vs. high, $p = .097$). A value of $p = 1.000$ reflects a Bonferroni-adjusted pairwise comparison, where adjusted values above 1 are capped at 1.000 by the statistical software. Bonferroni corrections were applied to the follow-up pairwise comparisons. 

\subsubsection{Correct Turns}

For analysis, we defined a correct first turn as the user's initial rotation (right or left) that takes the shortest path towards the target (right or left).
We analyzed whether haptic feedback influenced the likelihood of making a correct first turn using a Bayesian Beta--Binomial model with uniform Beta$(1,1)$ priors. 
Posterior estimates indicated that the probability of a correct turn was higher with haptic feedback ($M = 0.32$, $95\%$ CI $[0.28, 0.36]$) than without ($M = 0.24$, $95\%$ CI $[0.20, 0.28]$). 
The posterior distribution of the difference suggested that haptic feedback improved performance by about $8\%$ on average ($95\%$ CI $[0.02, 0.14]$), with a $99.7\%$ posterior probability that haptics were beneficial. 
These results provide strong evidence that haptic feedback helps participants initiate their turn in the correct direction.

\subsection{Qualitative Analysis}

Analysis of participant short-answer responses revealed two major themes. First, haptic feedback clearly improved horizontal localization, with 35 mentions across responses describing vibrations as making left–right orientation intuitive and reliable (e.g., \textit{``It can guide me to a correct direction''}; \textit{``Vibrations helped to get a general direction in which to look''}). 

Second, vertical guidance was more mixed and ambiguous, with 34 mentions spanning four subpatterns. Several participants described vertical cues as informative (e.g., \textit{``height mostly just told me where to look,''} \textit{``helpful for height,''} \textit{``When vibrations were present, I first identified the height at which the target existed, and then used the differences in vibration intensity to search for it,''} \textit{``When vibrations were present, the height was intuitively easy to understand, so I didn’t have to search up and down''}). Some participants reported difficulty distinguishing height or found it not helpful (e.g., \textit{``didn't determine the height,''} \textit{``bit difficult with the height difference,''} \textit{``height sensation was not intuitive,''} \textit{``harder than the recognition in the x-axis''}), while others defaulted to vision when judging vertical position (e.g., \textit{``mostly still looked up and down,''} \textit{``Regarding height, I was using visual information to align''}). A small group suggested that vertical feedback required more familiarization (e.g., \textit{``familiarizing over a longer period of time,''} \textit{``bit more training and exposure time,''} \textit{``felt it took some getting used to''}).

\subsection{Summary}

The results of our lab study indicate that haptic feedback significantly improved overall performance by reducing task duration compared to trials without haptics.
Furthermore, the presence of haptics increases users' confidence in localizing the target object.
While height alone did not have a reliable main effect, follow-up comparisons revealed that, in the absence of haptics, performance at the low height was significantly worse than at both the medium and high heights. In the combined haptic condition, all height levels performed comparably.
Similar effects were revealed for horizontal switches and target overshoots, while for vertical switches only height differences influenced behavior.
Moreover, haptics was shown to improve decision making as the presence of haptics significantly improved the correctness of the user's first turn.
Subjective impressions supported previous results, as participants indicated haptics clearly helped them in localizing targets.
On the other hand, height differences through frequency were not found to be intuitive, requiring more familiarization.

\section{Discussion}

The results of both studies suggest that tactile search can support visual search performance. The results of the lab study also provide insight into how users adapt their strategies and behaviors when using the method. The findings also imply that a spatial haptics method could be effective for the representation of height.

Across both studies, haptic feedback was associated with faster performance and higher confidence. This aligns with prior research showing search-related benefits of vibrotactile cues~\cite{Beese2025Feel, lehtinen_dynamic_2012, tivadar_digital_2022}.

The differences between Studies 1 and 2 in the strength of the haptic effect and the pattern of height-related effects could be explained by differences in sample size or setting, which may have affected how participants allocated attention during the task.

Beyond raw speed, participants’ behavior revealed that haptics supported more stable strategies: fewer unnecessary turns and overshoots indicated greater decisiveness, which in turn contributed to shorter search times. This behavioral pattern resonated with the qualitative reports. Participants generally agreed that haptics improved localization and direction, noting that the vibrations made it easier to find and orient toward targets: \textit{``The vibration helps me to locate the ball efficiently.''} and \textit{``It can guide me to a correct direction.''} The clearest benefit was in horizontal guidance, with many describing left–right localization as natural or even subconscious: \textit{``Horizontal made localization intuitive.''} and \textit{``Affected my targeting subconsciously.''} 

Our implementation combined horizontal amplitude cues with frequency cues for vertical information using two controllers. Without haptics, participants identified low-height targets more slowly than medium and high targets, whereas no differences between heights were observed in the combined haptic condition. Because the haptic condition combined horizontal and vertical cues, these results do not isolate the contribution of frequency modulation, and some of the outcomes could potentially be explained by the contributions of horizontal cues alone. This result is notable in light of eye-tracking research showing that people naturally fixate near the center of scenes~\cite{tatler_central_2007}, with a central fixation distribution that is typically narrower vertically than horizontally~\cite{clarke_deriving_2014}, and some evidence of an additional preference for the upper half of the screen during search~\cite{pflugshaupt_linking_2009}.

Interestingly in contrast to the horizontal dimension, height or vertical information was perceived to be less effective, with some finding it harder to notice or less helpful: \textit{``A bit difficult with the height difference''} and \textit{``More difference of vibration of directions than of the height.''} In these cases, vision was perceived to dominate vertical judgments, as reflected in comments such as \textit{``Easier with eyes for height''} and \textit{``the vibrations weren't enough to tell the height, so I had to rely mainly on my vision''}, and \textit{``Regarding height, I was using visual information to align.''} 

Possible reasons for why vertical was perceived as less effective include the asymmetry between bilateral continuous amplitude for horizontal (an interaural level difference analog from the auditory system) and monaural discrete frequency for vertical (a pitch-height crossmodal correspondence), both drawn from auditory perception, the first being a direct bilateral comparison and the second a learned crossmodal association. 

Beyond functionality and application, several described the game in positive, affective terms, calling it \textit{``fun''} and \textit{``creative.''}

\subsection{Future Directions and Limitations}
Future work should investigate long-term learning to understand how performance and experience with spatial and multidimensional mappings become more transparent and embodied with practice, as seen in sensory substitution adaptation and perceptual learning~\cite{maimon_case_2022, martolini2020shape, cuppone_audio-motor_2019, brandebusemeyer_impact_2020}. Examining interactions with crossmodal correspondences, such as the association of pitch with height~\cite{chiou2012cross}, is a promising direction to explore further, as previous results in multimodal and musical haptics are mixed, leaving an open question~\cite{occelli_compatibility_2009, aker_evaluation_nodate}. Additional opportunities include encoding signal identity and integrating multiple simultaneous haptic channels. Testing in naturalistic environments with clutter, competing stimuli, and systematically varied target velocities will be essential for assessing practical deployment. Depth encoding was not evaluated, although theoretically feasible.
We compared haptic guidance only with a vision only baseline, and interactions with spatial audio or multimodal cues remain unexplored. Participants engaged with the system only briefly, and qualitative responses suggest that extended familiarization may improve perception of vertical cues. Finally, tasks were performed in VR, which offers controlled measurement but does not fully represent real world complexity.

Because the present study compared continuous haptic guidance with no haptic guidance, it cannot determine whether feedback must remain active after initially orienting the user toward the target. The higher probability of a correct initial turn suggests that some benefit may occur early in the search, whereas the condition-specific differences in switching and overshooting are consistent with continued feedback supporting subsequent corrections. Future work should compare continuous guidance with brief initial cues and feedback that stops once the target enters the field of view.

Although observed power in experiment 2 was approximately .60, the study detected significant effects, indicating that the sample was sufficient to detect effects of the observed magnitude~\cite{Hoenig2001Abuse}. Future studies with larger samples should assess the stability of these findings.

A systematic evaluation of the framework in general, including comparison to a horizontal-only representation, and an evaluation of mental workload from different sensory cues in relation with performance could be very valuable. Furthermore, the lower perceived effectiveness of the vertical representation could be further studied using varying frequency encoding.

\subsection{Conclusion}

Spatial haptic feedback supports visual search in three dimensional environments, improving performance, confidence, and behavioral efficiency. The results further indicate that vertical position may be conveyed through vibrotactile cues and that doing so could mitigate asymmetries in search performance across heights. These findings suggest that vibrotactile spatial guidance can be effective for dynamic 3D targeting and indicate opportunities for deploying haptic guidance in ubiquitous computing scenarios where attention is divided and sensory resources are limited.

\acknowledgments{
During the preparation of this work, the authors used ChatGPT (OpenAI) and Claude (Anthropic) to improve the readability and language of the manuscript.

This work was supported in part by the German Research Foundation (DFG) through Collaborative Research Center (Sonderforschungsbereich) 1320, Project-ID 329551904, ``EASE---Everyday Activity Science and Engineering,'' and by the Minds, Media, Machines High-Profile Research Area at the University of Bremen (MMM Seed Grant No. 014). This research was also supported by the Marsden Fund under Grant MFP-UOC2516 and by the Cooperative Research Project Program of the Research Institute of Electrical Communication, Tohoku University.
}

\bibliographystyle{abbrv-doi}

\bibliography{main}
\end{document}